\documentclass[12pt]{article}
\usepackage{amsmath}
\usepackage{amsfonts}
\usepackage{graphicx}
\usepackage[capitalise]{cleveref}
\crefformat{equation}{(#2#1#3)}
\usepackage{xcolor,soul}
\usepackage[linesnumbered,ruled,vlined]{algorithm2e}

\usepackage{url}
\usepackage{makecell}
\usepackage{authblk}

\newcommand{\SA}{\texttt{SA}\xspace}
\newcommand{\QESA}{\texttt{QESA}\xspace}
\newcommand{\QA}{\texttt{QA}\xspace}
\newcommand{\Scipy}{\texttt{Scipy}\xspace} 
\newcommand{\Gurobi}{\texttt{Gurobi}\xspace} 
\providecommand{\keywords}[1]
{
	\small	
	\textbf{\textit{Keywords---}} #1
}
\usepackage{enumitem}
\setlist{topsep=3pt, itemsep=3pt, parsep=0pt}

\title{Extending quantum annealing to continuous domains: a hybrid method for \\quadratic programming}
\author{Hristo N.\ Djidjev}
\affil{Los Alamos National Laboratory\\ Los Alamos, NM 87545, USA \\
	and \\
	Institute of Information and Communication Technologies\\ Bulgarian Academy of Sciences, Sofia, Bulgaria 
	}

\begin{document}
	
	\maketitle
	\begin{abstract} 
		We propose Quantum Enhanced Simulated Annealing (QESA), a novel hybrid optimization framework that integrates quantum annealing (QA) into simulated annealing (SA) to tackle continuous optimization problems. While QA has shown promise in solving binary problems such as those expressed in Ising or QUBO form, its direct applicability to real-valued domains remains limited. QESA bridges this gap by using QA to select discrete search directions that guide SA through the continuous solution space, enabling the use of quantum resources without requiring full problem discretization. We demonstrate QESA's effectiveness on box-constrained quadratic programming (QP) problems, a class of non-convex optimization tasks that frequently arise in practice. Experimental results show that QESA consistently outperforms classical baselines in solution quality, particularly on larger and more ill-conditioned problems, while maintaining competitive runtime. As quantum annealing hardware matures, QESA offers a scalable and flexible strategy for leveraging quantum capabilities in continuous optimization.
	\end{abstract}
	
	
	\section{Introduction}
		
	Quantum annealing (QA) is a metaheuristic designed to solve combinatorial optimization problems by exploiting quantum mechanical phenomena such as tunneling and superposition~\cite{das2008colloquium,kadowaki1998quantum,Morita_2008}. It is particularly effective for problems that can be formulated as minimizing the energy function of an Ising model or, equivalently, as solving a quadratic unconstrained binary optimization (QUBO) problem~\cite{lucas2014ising}. Examples of such problems include graph coloring~\cite{kwok2020graph}, the maximum clique~\cite{chapuis2017finding}, and the minimum vertex cover~\cite{pelofske2019solving}. These binary formulations are well-suited to current QA hardware, such as the D-Wave systems, which are designed to natively operate on binary spin variables. QA is inspired by the framework of adiabatic quantum computation~\cite{RevModPhys.90.015002,farhi2000quantum}, in which a quantum system is slowly evolved from an easily prepared initial ground state to the ground state of a Hamiltonian that encodes the objective function. However, current hardware implementations deviate from this idealized model, operating instead in an open-system, non-adiabatic regime.
	
	Despite its effectiveness on binary problems, quantum annealing cannot be directly applied to continuous or non-binary optimization tasks, which are common in many real-world domains such as engineering design, resource allocation, and finance. The binary constraint imposed by QA hardware requires problem reformulations that encode continuous variables by binary ones. 
	
	While quantum annealing is well-suited for binary optimization problems, its extension to continuous or mixed-variable domains remains relatively underexplored. One direction of research involves encoding each continuous or integer variable using a set of binary variables—typically via unary, domain-wall, or fixed-point encodings—so that the resulting formulation can be mapped to a QUBO problem solvable by QA. For example, Arai et al.~\cite{arai2023effectiveness} applied domain-wall encoding to find the minimum a continuous single-variable function by mapping it to an Ising model and benchmarked performance on a D-Wave annealer. Chang et al.~\cite{chang2020integer} developed a method for solving integer linear programs using bit-level encodings of integer variables. Similarly, Iftakher et al.~\cite{iftakher2023mixed} proposed a QUBO-based formulation for solving mixed-integer quadratic programs using binary and unary encodings for both continuous and integer variables.
	
	An alternative approach is to design hybrid schemes that incorporate quantum annealing into broader classical workflows. Ottaviani and Amendola~\cite{ottaviani2018lowranknonnegativematrix} explored a quantum-classical algorithm using reverse annealing for low-rank matrix factorization problems with real-valued variables  and applied it to a $2\times 2$ matrix. This method leverages quantum refinement starting from classical initial guesses to explore difficult nonconvex landscapes. 
	
	Although these studies demonstrate the potential of quantum annealing for problems involving continuous or integer variables, most existing methods suffer from significant limitations. Representing a single continuous or integer variable using multiple binary variables introduces encoding overhead that severely restricts the problem sizes that current QA hardware can handle. In addition, the resulting formulations often contain coefficients of widely varying magnitudes, which can lead to ill-conditioned energy landscapes that degrade solution quality. Many of these approaches also rely on substantial classical preprocessing or postprocessing. As a result, a coherent and scalable framework for applying QA to continuous optimization remains underdeveloped—especially when contrasted with the maturity of classical methods—and motivates the exploration of new hybrid strategies that extend the reach of QA to non-binary domains.
	
	In this work, we propose a novel hybrid optimization method for continuous problems that embeds QA into a simulated annealing (SA) loop. Specifically, we use QA to determine a promising descent direction at each SA iteration by solving an Ising-encoded subproblem. The resulting direction is used to update a continuous solution vector, with feasibility ensured via simple projection (clipping). We apply this framework to the important class of quadratic programming (QP) problems with box constraints, where the objective is to minimize a quadratic function over a bounded domain.
	
	To our knowledge, our method is the first to use quantum annealing specifically for selecting descent directions in a continuous optimization framework. It combines the efficiency of QA in identifying near-optimal directions from an exponentially large discrete search space with the adaptability of simulated annealing to explore the solution space at multiple scales. This enables effective navigation of high-dimensional, non-convex landscapes without relying on discretization of variables or encoding schemes.

	\paragraph{Our contributions.} The main contributions of this work are as follows:
	\begin{itemize}
		\item We propose a hybrid QA-SA optimization framework for solving continuous quadratic programs with box constraints.
		\item We introduce a novel use of quantum annealing to determine discrete descent directions in continuous space.
		\item We evaluate our method on randomly generated QP instances, demonstrating that it can match or exceed the quality of classical solvers like Gurobi and Scipy.
		\item We analyze key components of our method, including the effects of boundary-based initialization, iterations step count, and the quality of QA-derived directions.
	\end{itemize}
	
	The rest of the paper is organized as follows. Section~\ref{sec:methods} provides background on quantum annealing, Ising models, and classical simulated annealing and introduces the QESA framework, including its directional optimization strategy, initialization scheme, and integration of quantum and classical components. In Section~\ref{sec:results}, we present an extensive experimental evaluation of QESA on box-constrained quadratic programming problems, comparing its performance against classical and quantum baselines. Finally, Section~\ref{sec:conclusion} summarizes our findings and outlines potential directions for future research.

	\section{Methods}\label{sec:methods} 
	
	\subsection{Quantum annealing and Ising models }	
	Quantum annealing (QA) is a physics-inspired optimization technique that leverages quantum fluctuations to explore the solution space of difficult combinatorial problems \cite{farhi2000quantum, kadowaki1998quantum}. It is particularly well-suited for minimizing objective functions expressed as Ising models, which represent a class of energy functions defined over binary spin variables.
	
	The classical Ising model is defined by an energy function over \( n \) binary spin variables \( s_i \in \{-1, +1\} \), typically written as:
	\begin{equation}
			E(s) = \sum_{i < j} J_{ij} s_i s_j + \sum_i h_i s_i.
	\end{equation}

	Here, \( J_{ij} \) represents the coupling strength between spins \( s_i \) and \( s_j \), while \( h_i \) denotes the local field acting on spin \( s_i \). The goal is to find the spin configuration \( s = (s_1, \dots, s_n) \) that minimizes the energy function \( E(s) \).
	
	QA solves this problem by initializing the system in an equal superposition of all spin states and gradually reduces quantum fluctuations according to a predefined annealing schedule. The process exploits quantum tunneling to escape local minima and ideally converges to a global minimum of the energy landscape. The final spin configuration obtained at the end of the anneal corresponds to a candidate solution to the original optimization problem.
	
	Modern QA hardware, such as the D-Wave quantum annealers, natively support Ising model optimization by embedding the problem into a physical quantum system. The user specifies the \( J_{ij} \) and \( h_i \) parameters, and the annealer returns low-energy spin configurations sampled from the problem’s energy landscape. While solution quality may depend on hardware noise and annealing parameters, QA has shown promise in efficiently finding good-quality solutions for large, rugged energy landscapes.
	
	Minimizing a quadratic function over binary variables is an NP-hard problem~\cite{Barahona1982}, and many important NP-hard problems—including maximum clique, max-cut, and vertex cover—can be naturally formulated as simple QUBO or Ising models~\cite{Lucas2014}. However, practical limitations such as limited qubit connectivity, analog control errors, and decoherence still present challenges to the widespread application of QA~\cite{king2022coherent, pearson2019analog, D-Wave-ICE}. Techniques such as minor embedding~\cite{choi2011minor}, quantum error correction~\cite{pudenz2014error}, and hybrid quantum-classical methods~\cite{djidjev2024enhancing} have been developed to mitigate these issues.
	
	In our approach, we harness QA not to solve the original continuous optimization problem directly, but rather to guide a classical search method by solving carefully constructed Ising subproblems. This enables the use of quantum annealing to handle discrete components within a continuous optimization framework, leveraging its combinatorial power while preserving scalability in real-valued domains.	
	
	\subsection{Quadratic programming}
	
	\textit{Quadratic programming} (QP) \cite{nocedal2006quadratic} refers to a class of optimization problems in which the objective function is quadratic and the feasible region is defined by linear constraints. In its simplest form, a QP can be written as: 
	\begin{equation}
			\min_{x \in \mathbb{R}^n} \quad \frac{1}{2} x^T Q x + c^T x \quad \text{subject to} \quad Ax \leq b \label{eq:qp}
	\end{equation}

	where \( Q \in \mathbb{R}^{n \times n} \) is a symmetric matrix, \( c \in \mathbb{R}^n \) is a linear coefficient vector, and \( A \in \mathbb{R}^{m \times n}, b \in \mathbb{R}^m \) define the linear inequality constraints.
		
	In this work, we focus on the case of QPs with bound constraints, also known as \textit{quadratic programming with box constraints}~\cite{burer2009nonconvex, DeAngelis1997}. Such constraints take the form:
	\[
	l_i \leq x_i \leq u_i \quad \text{for} \quad i = 1, \dots, n,
	\]
	where \( l_i \) and \( u_i \) are the lower and upper bounds on the \( i \)-th variable. This formulation arises frequently in practical applications such as portfolio optimization, resource allocation, and control, and is considered a fundamental problem in global optimization~\cite{burer2009nonconvex}.
	
	While QPs with box constraints can be solved efficiently using classical convex optimization techniques when \( Q \) is positive semidefinite, they become significantly more challenging when the matrix \( Q \) is indefinite, rendering the problem non-convex. In fact, solving box-constrained QPs in the general (non-convex) case is NP-hard~\cite{pardalos1991quadratic, sahni1974computationally}. As a result, heuristic, local search, or global optimization strategies are typically employed in practice.
	
	These types of problems are not directly amenable to quantum annealing, since QA is inherently designed to optimize functions of binary variables. Mapping a continuous variable to a binary spin system typically involves discretization or encoding schemes that introduce approximation errors and increase the size of the problem. Moreover, preserving constraint feasibility during such encoding is non-trivial. Instead of discretizing the entire QP, our approach uses quantum annealing only to select promising directions, which are then explored using simulated annealing as a flexible global search framework.

	\subsection{Simulated annealing }
	
	Simulated annealing (SA) is a stochastic optimization algorithm inspired by the physical process of annealing in metallurgy, where a material is heated and slowly cooled to minimize its internal energy \cite{kirkpatrick1983optimization,van1987simulated}. In optimization, SA is used to approximate the global minimum of an objective function by exploring the search space probabilistically.
	
	At each iteration, SA perturbs the current solution to generate a new candidate solution. If the candidate yields a lower objective value, it is accepted unconditionally. If it results in a higher objective value, it may still be accepted with a probability that decreases over time, governed by a "temperature" parameter \( T \). This probabilistic acceptance criterion allows SA to escape local minima and explore the solution space without being confined to local basins in the early stages.
	
	The acceptance probability \( P \) of a worse solution is typically given by the Metropolis criterion:
	\[
	P = \exp\left(-\frac{\Delta E}{T}\right),
	\]
	where \( \Delta E \) is the increase in the objective value and \( T \) is the current temperature. As the algorithm proceeds, the temperature is gradually reduced according to a predefined cooling schedule, leading to increasingly greedy behavior and convergence toward a minimum within the current basin of attraction.
	
	SA is particularly appealing for non-convex or combinatorial optimization problems where the solution space may contain many local minima. Its simplicity and robustness make it a valuable tool for black-box optimization tasks where gradient information is unavailable or unreliable.
	
	However, SA often suffers from slow convergence, especially in high-dimensional spaces or rugged energy landscapes. Its performance is highly sensitive to the choice of the cooling schedule, initial temperature, and neighborhood structure used to generate candidate solutions.
	
	In our hybrid approach, simulated annealing serves as the outer optimization framework, while quantum annealing is used to select promising descent directions at each step. This choice of search direction is formulated as an Ising model, as discussed in the next subsection.
	
	\subsection{Optimizing direction as an Ising  problem}
	
	At each iteration of simulated annealing, the current solution is refined by stepping in a discrete direction \( s \in \{-1, 1\}^n \), scaled by a  step size \( k \in \mathbb{R} \) that decreases with the iteration number. The update rule is given by:
	\[
	x_{\text{new}} = x + k \cdot s.
	\]
	
	To guide the search efficiently, we aim to select a direction \( s \) that minimizes the objective function when taking a step from the current point. Substituting this update rule into the objective function:
	\[
	f(x) = \frac{1}{2} x^T Q x + c^T x,
	\]
	we obtain:
	\[
	f(x + k \cdot s) = \frac{1}{2} (x + k s)^T Q (x + k s) + c^T (x + k s).
	\]
	Expanding and simplifying:
	\[
	f(x + k \cdot s) = f(x) + k (x^T Q + c^T)s + \frac{1}{2} k^2 s^T Q s.
	\]
	
	Because \( f(x) \) is constant for a given iteration, the direction selection reduces to minimizing the terms that depend on \( s \):
	\[
	\min_{s \in \{-1, 1\}^n} \quad \frac{1}{2} k^2 s^T Q s + k (x^T Q + c^T)s.
	\]
	
	This optimization objective aligns with the standard Ising model form
	\[
	\mathit{Is}(s) = s^T J s + h^T s,
	\]
	where \( J = \frac{1}{2} k^2 Q \) and \( h = k (Q x + c) \). Thus, finding the optimal direction \( s \) corresponds to finding the ground state of an Ising system, a task that is well-suited for quantum annealing.
	
	In addition to guiding search directions during SA optimization, quantum annealing is also used in our framework to generate high-quality initial states, as described in the next subsection.
	
	\subsection{Initial state selection using QA}
	
	To improve the performance of the algorithm, we initialize optimization from a corner of the feasible region by choosing an initial vector \( x \in \{-1, 1\}^n \), corresponding to the corners of the hypercube defined by the box constraints \( x_i \in [-1,1] \). 
	This choice is motivated by empirical observations: initializing on the boundary often leads to better convergence and improved solution quality, particularly in high-dimensional settings, as further examined in our results section.
	
	To find such an initial point, we solve an Ising model whose ground state corresponds to a good candidate among the corner configurations. This is done by restricting the original QP objective from \cref{eq:qp} to binary variables and minimizing the resulting function
	\[
	\min_{x \in \{-1,1\}^n} \Big(\;\frac{1}{2} x^\top Q x + c^\top x\;\Big),
	\]
	where \( Q \in \mathbb{R}^{n \times n} \) is symmetric and \( c \in \mathbb{R}^n \). Our goal is to reformulate this binary QP into the standard Ising model form:
	\[
	H_{\text{Ising}}(x) = \sum_{i < j} J_{ij} x_i x_j + \sum_i h_i x_i.
	\]
		
	Starting with:
	\[
	\frac{1}{2} x^\top Q x + c^\top x = \frac{1}{2} \sum_{i,j} Q_{ij} x_i x_j + \sum_i c_i x_i,
	\]
	we separate the terms:
	\[
	= \frac{1}{2} \sum_{i \neq j} Q_{ij} x_i x_j + \frac{1}{2} \sum_i Q_{ii}x_i^2 + \sum_i c_i x_i.
	\]
	Because \( x_i^2 = 1 \) for \( x_i \in \{-1,1\} \), we note that \( \frac{1}{2} \sum_i Q_{ii} \) is a constant offset that does not depend on \( x \) and can be therefore dropped from the objective.
	
	Combining terms and symmetrizing, we get
	\[
	H_{\text{Ising}}(x) = \sum_{i < j} Q_{ij} x_i x_j + \sum_i c_i x_i,
	\]
	which allows us to determine the Ising model parameters:
	\[
	J_{ij} = Q_{ij}, \quad h_i = c_i.
	\]
	
	\subsection{Putting all elements together: the QESA framework}
	
	We now present our hybrid optimization framework, called \textit{quantum enhanced simulated annealing (QESA)}, which integrates SA with QA for solving box-constrained quadratic programs. In summary, QESA performs a series of quantum-guided updates within a simulated annealing loop. Quantum annealing proposes efficient directions, while simulated annealing ensures exploration and feasibility in the continuous domain. The combined method balances local refinement and global search, with constraint feasibility enforced by projection. The overall procedure is summarized in Algorithm~\ref{alg:pseudocode}.
	
	Each iteration of the algorithm involves the following steps:
	\begin{enumerate}
		\item \textbf{Direction selection:} Use quantum annealing to solve an Ising model that identifies a descent direction \( s \in \{-1,1\}^n \).
		\item \textbf{Step update:} Compute a candidate point using the update rule \( x' = x + k \cdot s \), where \( k \) is a step size, typically reduced at each iteration.
		\item \textbf{Acceptance:} Use the Metropolis criterion to decide whether to accept the candidate \( x' \).
	\end{enumerate}
	
	The algorithm proceeds iteratively according to the steps above until a termination condition is met. Termination can be based on a fixed number of iterations, a temperature threshold, or lack of improvement in the objective value over time.
	
	To ensure feasibility with respect to the box constraints, we apply \emph{element-wise clipping}:
	\[
	x'_i \leftarrow \min(\max(x_i + k s_i, -1), 1).
	\]
	This guarantees that the updated point remains in the valid domain \( [-1,1]^n \). As a result, the box constraints are enforced implicitly, allowing us to use the original quadratic objective without introducing penalty terms into the energy function.
	
	\begin{algorithm}
		\DontPrintSemicolon
		\KwIn{Matrix \( Q \), vector \( c \), initial step size \( k \), step scaling factor \( \alpha \), initial temperature \( T_0 \), cooling schedule}
		\KwOut{Optimized vector \( x \in [-1,1]^n \)}
		Assume domain \( [-1,1]^n \); rescale inputs if necessary\;
		Solve Ising problem to obtain initial state \( x \in \{-1,1\}^n \)\;
		\While{termination criterion not met}{
			Construct Ising model: \
			\hspace{1em} \( H(s) = \frac{1}{2} k^2 s^T Q s + k (Qx + c)^T s \)\;
			Solve for \( s \in \{-1,1\}^n \) using quantum annealing\;
			Propose new point: \( x' \leftarrow x + k \cdot s \)\;
			Clip: \( x'_i \leftarrow \min(\max(x'_i, -1), 1) \) for all \( i \)\;
			Compute \( \Delta f = f(x') - f(x) \)\;
			Accept \( x' \) with probability \( \min(1, \exp(-\Delta f / T)) \)\;
			\If{accepted}{
				\( x \leftarrow x' \)\;
			}
			Update step size \( k \leftarrow k\alpha \)\;
			Update temperature \( T \leftarrow \text{schedule}(T) \)\;
		}
		\Return \( x \)\;
		\caption{Quantum Enhanced Simulated Annealing (QESA) algorithm.}\label{alg:pseudocode}
	\end{algorithm}
	
	This framework leverages quantum annealing to guide local search while preserving the flexibility and global reach of simulated annealing.		
	
	\section{Results}\label{sec:results} 
	
	\subsection{Experimental setup}
	
	We implemented our QESA framework in Python and conducted experiments using D-Wave's  \texttt{Advantage\_system4.1} quantum annealer, accessed via the Leap quantum cloud platform. Quantum annealing parameters were set to \texttt{num\_reads} = 1000 and \texttt{annealing\_time} = 100 microseconds, consistent with typical settings used in prior work to balance performance and runtime. All other QA parameters were left at their default values.
	
	To evaluate performance, we generated random QP instances with varying problem sizes and conditioning. Specifically, we considered problem dimensions \( n \in \{50, 100, 150\} \) and used symmetric \( Q \) matrices with adjustable diagonal dominance. All entries of \( Q \), both diagonal and off-diagonal, were initially sampled uniformly at random from the interval \( [-1, 1] \). To control diagonal dominance, the diagonal entries were then scaled by a factor selected from the set \( \{1, 5, 10, 20\} \), while the off-diagonal entries remained unchanged. The resulting matrix was dense with a fill ratio (density) \( d = 1 \).
	For each combination of matrix generation parameters, we generated 5 random instances.
	
	Diagonal dominance affects the curvature of the quadratic objective function and influences whether optimal solutions lie on the boundary or in the interior. Varying the diagonal scale allows us to systematically influence this likelihood. In the next subsection, we analyze how the diagonal scaling factor affects the distribution of solution values relative to the domain boundary. 
	
	We compared five algorithms:
	\begin{enumerate}
		\item \QESA: implementation of our proposed hybrid framework that uses quantum annealing to guide search directions within simulated annealing.		
		\item \QA: quantum annealing applied to the binary formulation of the QP problem as a standalone algorithm. In \QESA, \QA is used to compute an initial solution. Note that we use \SA/\QA to refer to the specific implementations evaluated in our experiments, while SA/QA refer to the general optimization methods.		
		\item \SA: a classical simulated annealing method with the same structure as \QESA, but using SA instead of QA for both the initial solution and direction selection.		
		\item \Scipy: the \texttt{SLSQP} (sequential least squares programming) method from the \texttt{scipy.optimize} Python module with default options. The \texttt{scipy} library \cite{2020SciPy-NMeth} is widely used in scientific computing and provides a convenient baseline for continuous optimization tasks.
		\item \Gurobi: a commercial state-of-the-art QP solver, used as a benchmark. It is widely regarded as one of the most powerful optimization solvers and can certify global optimality for non-convex QP problems when solved to completion  \cite{gurobi}.
	\end{enumerate}
	
	\begin{table*}[!t]
		\renewcommand{\arraystretch}{1.5}
		\centering
		\begin{tabular}{|c|c|c|c|}
			\hline
			Name     & Optimization Method            & \makecell{Optimality\\guarantee} & \makecell{Classical/\\Quantum} \\ \hline\hline
			\texttt{QESA}     & hybrid SA with QA      & No                          & hybrid            \\ \hline
			\texttt{QA}       & quantum annealing      & No                          & quantum           \\ \hline
			\texttt{SA}       & simulated annealing    & No                          & classical         \\ \hline
			\texttt{Scipy}    & sequential least squares (SLSQP) & No                   & classical         \\ \hline
			\texttt{Gurobi}   & branch-and-cut         & Yes                         & classical         \\ \hline
		\end{tabular}
		\caption{Summary of algorithms used in the experimental evaluation.}
		\label{tab:alg_summary}
	\end{table*}	
	
	The simulated annealing component was implemented using a custom wrapper based on the \texttt{simanneal} package \cite{simanneal}, modified to include quantum-enhanced steps when appropriate. The \SA and \QESA schedules used logarithmic cooling from \( T_{\max} = 1000 \) to \( T_{\min} = 0.1 \), with 100 steps. The initial step size and the step-size scaling factor $\alpha$ for \QESA are set to 0.1 and 0.95, respectively.
	
	Each algorithm was evaluated on every (dimension, diagonal scale, seed) combination, resulting in a comprehensive comparison across 60 different problem instances. The solution, objective value, convergence behavior, and computation time were recorded for all methods. Timing measurements include wall-clock time, with \QESA also reporting cumulative quantum processing unit (QPU) time.
	
	The next sections present detailed analysis of solution distributions, error trends, execution times, and the impact of iteration limits on algorithm performance.
	
	\subsection{Solution distribution analysis}
	\label{sec:distr}
	
	Since our hybrid QESA framework initializes the search from a point on the boundary of the feasible box—specifically, where each coordinate \( x_i \in \{-1,1\} \)—it is important to assess whether optimal or near-optimal solutions are more likely to lie near the boundary or within the interior of the domain. To investigate this, we analyzed the empirical distribution of solution coordinates across a wide range of problem instances.
	
	One structural factor that may influence this distribution is the relative dominance of diagonal entries in the matrix \( Q \). To study this effect, we grouped problem instances based on the diagonal scaling factor used when generating \( Q \), choosing values from the set \( \{1, 5, 10, 20\} \). For each group, we collected solution vectors returned by the Gurobi solver across all seeds and problem sizes, and generated histograms of the individual coordinate values. Gurobi was run with a time limit of one hour, meaning that the returned solution is either globally optimal or represent the best one found within the allotted time.
	
\begin{figure}
		\centering
		\includegraphics[width=\textwidth]{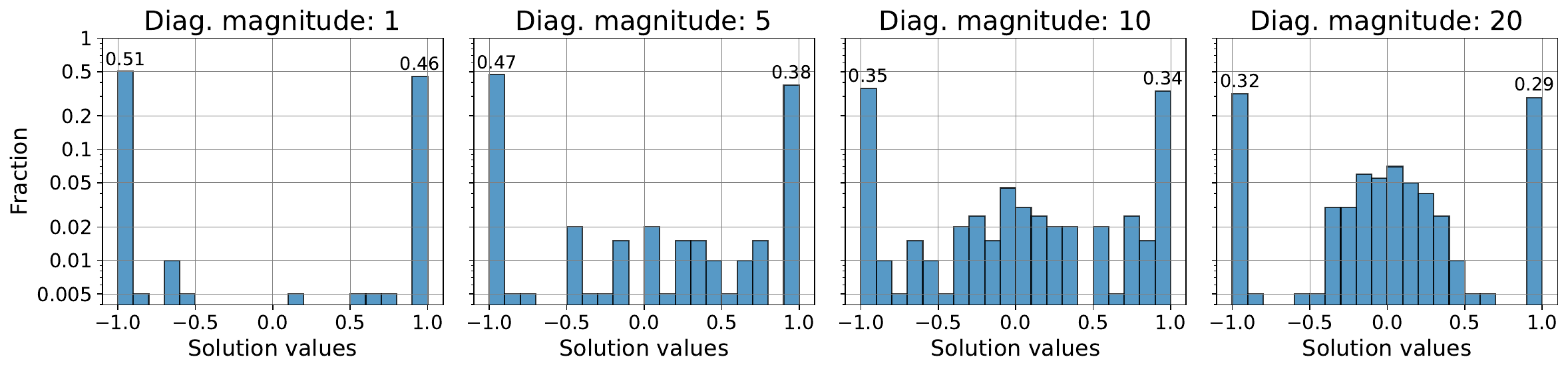}
		\caption{Distribution of solution values for \QESA solutions, grouped by diagonal scale. Each subplot corresponds to a different diagonal scaling factor applied to the matrix \( Q \). The y-axis uses a logarithmic scale to highlight both dominant and rare value frequencies.}
		\label{fig:histograms}
	\end{figure}

	Figure~\ref{fig:histograms} shows the resulting distributions. When the diagonal magnitude is small (e.g., 1), solution coordinates are highly concentrated at the box boundaries \( \pm 1 \), with 97\% of values taking on one of these extremes. As the diagonal dominance increases (e.g., scale factors of 5, 10, or 20), the distributions become more dispersed and shift gradually toward the interior of the domain \([-1,1]\). Nonetheless, even at the highest scaling factor tested (20), 61\% of coordinates remain on the boundary.
	
	These findings empirically support our design choice to initialize QESA from a boundary point obtained via quantum annealing. They also suggest that the diagonal scale factor in \( Q \) plays a meaningful role in controlling whether solutions tend to lie on the boundary or interior, which is relevant for both initialization and direction selection strategies in hybrid optimization.
	
	\subsection{Solution quality trends across solvers}
	
	To assess the relative performance of different optimization methods, we analyzed the final objective values (energies) returned by each solver across a wide range of problem instances. \Gurobi, being a commercial-grade solver capable of producing provably optimal solutions given sufficient time, serves as our reference baseline. However, due to the large number of test cases, we imposed a 10-minute time limit per instance. As a result, while some \Gurobi solutions are optimal, many are near-optimal.
	
\begin{figure}
		\centering
		\includegraphics[width=\textwidth]{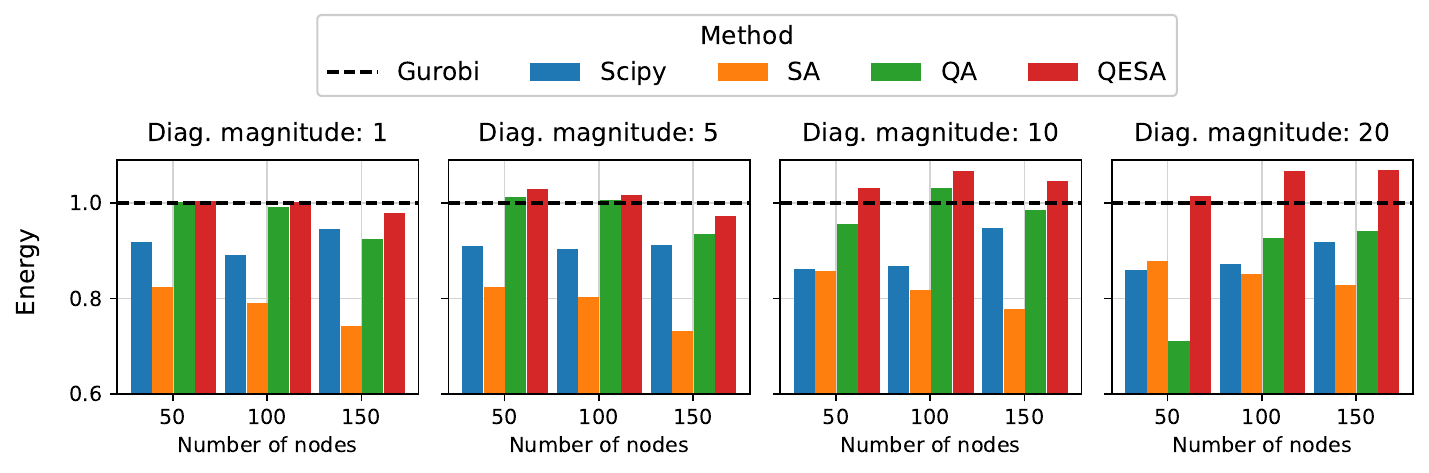}
		\caption{Relative energy error for each method across different problem sizes and diagonal scales. Each subplot corresponds to a different diagonal scaling factor applied to the QP matrix. Energies are normalized by the corresponding \Gurobi value.}
		\label{fig:facet-errors}
	\end{figure}

	For each method and instance type, we computed the energy of the final solution. Figure~\ref{fig:facet-errors} presents the results, with energies normalized by the corresponding \Gurobi energy for each instance, so that \texttt{Gurobi} always has a relative value of 1.
	
	Several performance patterns emerge:
	
	\begin{itemize}
		\item {\SA} yields the least accurate results overall. It performs reasonably well for small instances (\( n = 50 \)) but its errors increase significantly with problem size, especially when the diagonal dominance is low.
		
		\item {\Scipy} shows stable performance across all settings and problem sizes. However, its solution quality is generally lower than that of {\QA} and always lower than {\QESA}.
		
		\item {\QA} performs surprisingly well given its simplicity. In small-scale problems with low diagonal dominance, it often approaches the quality achieved by {\QESA}.
		
		\item {\QESA} consistently delivers the best performance among all methods, closely tracking or even outperforming \Gurobi in difficult cases. Notably, for large problem sizes (\( n \in \{100, 150\} \)) with strong diagonal scaling (\( \{10, 20\} \)), {\QESA} achieves lower-energy solutions than \Gurobi within the given time limit.
	\end{itemize}
	
	These results highlight \QESA’s ability to effectively combine the strengths of quantum annealing and simulated annealing. Methods that leverage quantum search components (\QA and \QESA) generally outperform their purely classical counterparts (\SA and \Scipy). These findings validate the benefit of incorporating quantum-enhanced direction selection into a classical optimization framework.
	
	\subsection{Execution time analysis}
	
\begin{figure}
		\centering
		\includegraphics[width=0.7\textwidth]{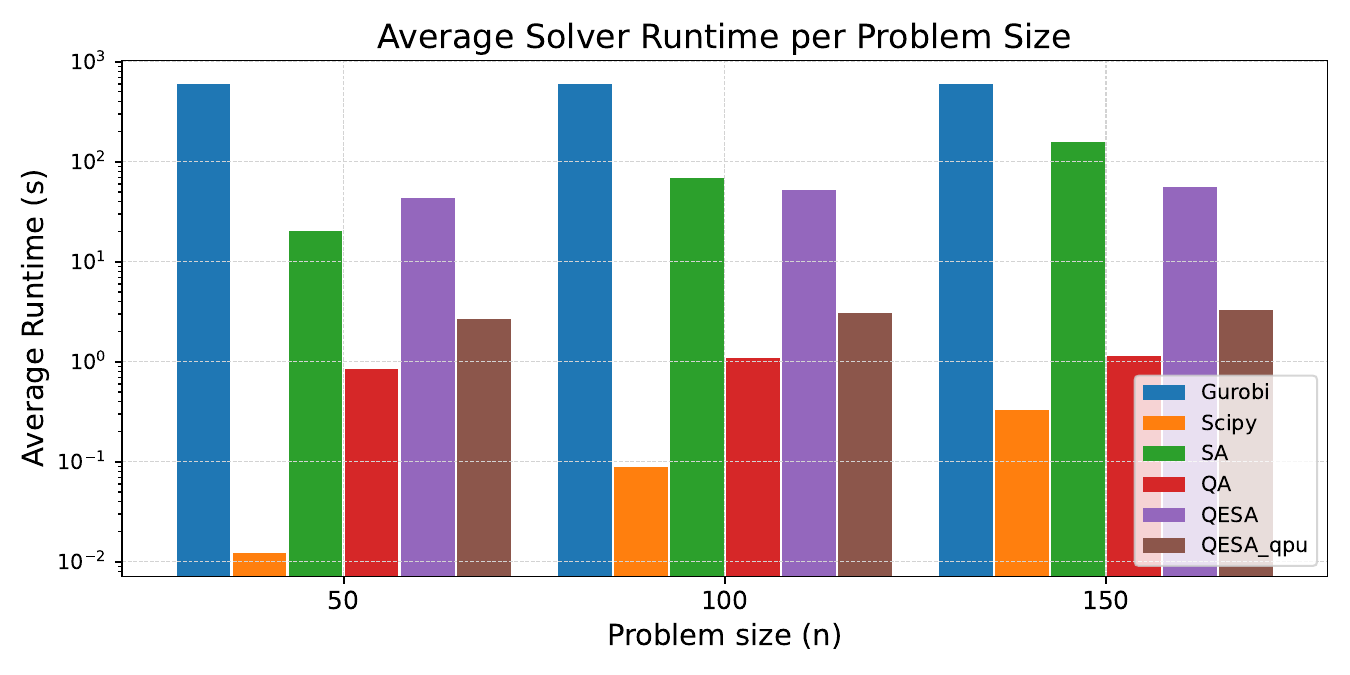}
		\caption{Average runtime per solver across different problem sizes (log scale on the y-axis). \texttt{QESA\_qpu} reports cumulative quantum processing time, while all other methods reflect wall-clock time.}
		\label{fig:avg-times}
	\end{figure}

	While the primary focus of this work is on solution quality within reasonable computational time, examining runtime behavior offers additional insights into the practicality of the methods. Figure~\ref{fig:avg-times} shows the average runtime for each solver across problem sizes \( n \in \{50, 100, 150\} \). The y-axis uses a logarithmic scale to account for the wide range of observed runtimes.
	
	The runtime for {\Gurobi} is fixed at 600 seconds for all instances due to the imposed time limit. In contrast, the other solvers exhibit varying behavior depending on algorithmic design and problem size.
	
	{\Scipy} is the most time-efficient method overall, consistently finishing in under a second. Its runtime grows slightly with \( n \), in line with expectations for classical gradient-based methods.
	
	The two quantum-assisted methods, {\QA} and {\QESA}, maintain nearly constant average runtimes regardless of problem size. This is due to the fixed number of iterations and the use of quantum annealing calls with constant annealing time (100 microseconds per call). These constant runtimes are made possible, in part, by the parallelism of quantum hardware, which allows thousands of qubits and couplers to operate simultaneously.
	
	However, the wall-clock times for {\QA} and {\QESA} include overhead from submitting jobs to D-Wave’s Leap cloud platform, including delays due to network latency and job queueing. As a result, their observed runtimes are significantly longer than the actual quantum processing time.
	
	To separate these effects, we also report the cumulative quantum processing time for {\QESA} under the label \texttt{QESA\_qpu}. These values represent the total time spent executing anneals on the QPU and remain on the order of a few seconds—even for the largest problem sizes.
	
	Overall, {\QESA} maintains a moderate computational footprint, with most of the runtime stemming from communication overhead rather than quantum computation. Among classical methods, {\Scipy} remains the fastest, making it an attractive choice when execution speed is prioritized over accuracy.
	
	\subsection{Effect of iteration count on solution quality}
	
	A key design parameter in {\QESA} is the iteration count, which balances computational effort with solution accuracy. To evaluate this impact, we ran {\QESA} with a step counts in \(\{5, 10, 20, 40, 60, 80, 100\}\) and measured the final energy achieved for problem sizes \(n \in \{50, 100, 150\}\). For comparison, all energies are normalized by the energy returned by {\Gurobi} with a one-hour timeout, which defines the baseline at \( y = 1 \).
	
\begin{figure}
		\centering
		\includegraphics[width=\textwidth]{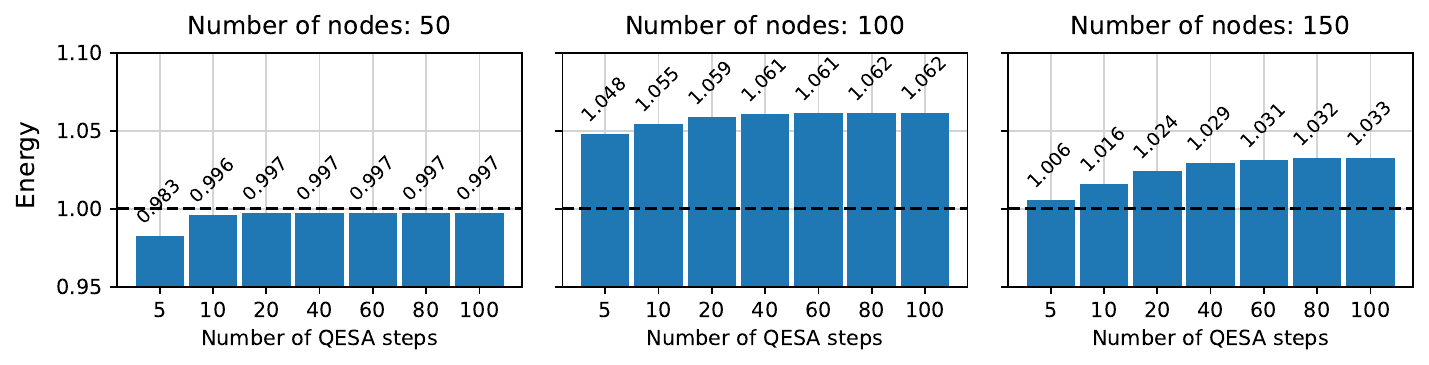}
		\caption{Normalized final energy as a function of the number of \QESA iterations, for problem sizes \(n = 50, 100, 150\). Energies are normalized relative to the \Gurobi solution. Each bar corresponds to a full \QESA run using a temperature schedule matched to the given number of steps.}
		\label{fig:qesa-steps}
	\end{figure}

	Figure~\ref{fig:qesa-steps} presents the results. Each datapoint represents a complete {\QESA} run using a cooling schedule scaled to the given step count, not an intermediate snapshot. We observe the following trends:
	
	\begin{itemize}
		\item For small problems (\(n = 50\)), near-optimal solutions are often found with as few as 20 steps. Additional iterations yield only minor improvements.
		\item For moderate problems (\(n = 100\)), convergence improves steadily up to about 60 steps, after which the gains diminish.
		\item For large problems (\(n = 150\)), the benefits of additional steps remain more significant, but performance tends to stabilize around 80–100 steps.
	\end{itemize}
	
	These results show that {\QESA} can produce high-quality solutions even with a modest number of iterations. Notably, for larger problem instances (\(n = 100\) and \(n = 150\)), {\QESA} consistently achieved lower-energy solutions than {\Gurobi}, despite the latter being given significantly more time.
	
	\subsection{Effect of boundary initialization on performance}
	
	\QESA is initialized by solving an Ising model that restricts the initial point to lie on the boundary of the feasible domain—that is, each coordinate \( x_i \in \{-1, 1\} \). To test whether this boundary-based initialization heuristic contributes meaningfully to solution quality, we designed an experiment in which the direction vector \( s \in \{-1,1\}^n \) computed by QA is partially replaced with random real values in the interval \( [-1, 1] \).
	
	We define a probability parameter \( p \in [0,1] \) that controls this replacement:
	\begin{itemize}
		\item When \( p = 1 \), the original QA-derived direction is used unchanged (i.e., fully discrete and boundary-aligned).
		\item When \( p = 0 \), the direction vector is completely randomized within the bounds.
		\item Intermediate values of \( p \) result in a hybrid direction vector, where each entry has a \( p \) chance of being kept and a \( 1 - p \) chance of being replaced by a random continuous value in \( [-1, 1] \).
	\end{itemize}
	
\begin{figure}
		\centering
		\includegraphics[width=\textwidth]{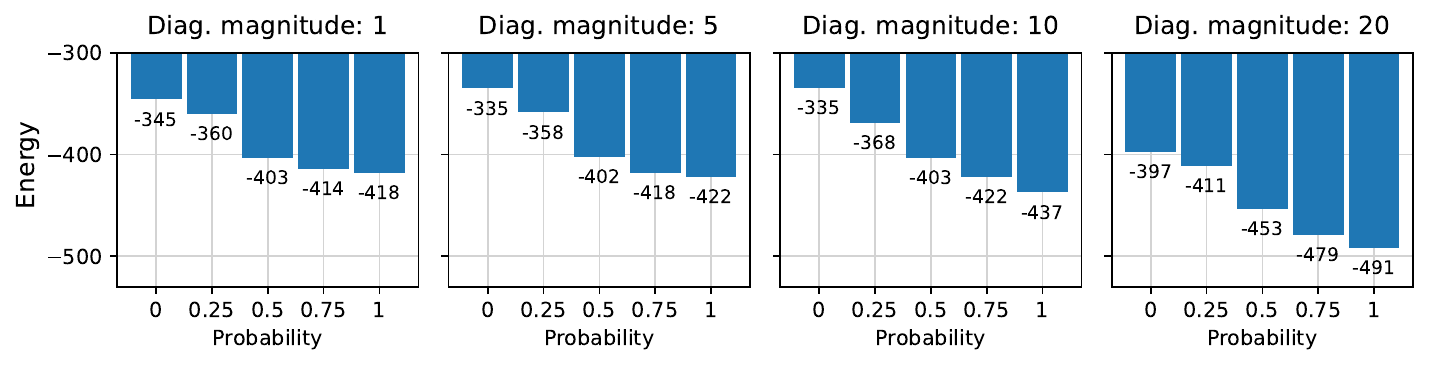}
		\caption{Final \QESA energy as a function of the probability \( p \) of retaining each component of the QA-computed direction vector. Results are grouped by diagonal magnitude of the Q matrix. Lower energy indicates better performance.}
		\label{fig:init-boundary}
	\end{figure}

	Figure~\ref{fig:init-boundary} shows the final energy values achieved by \QESA under varying values of \( p \), grouped by the diagonal dominance of the \( Q \) matrix. The following patterns emerge:
	
	\begin{itemize}
		\item For all problem instances, the best results are achieved when \( p = 1 \), meaning the initial state is in \( \{-1,1\}^n \). This supports our design choice in \QESA.
		\item Performance consistently degrades as \( p \) decreases, across all diagonal scaling factors.
	\end{itemize}
	
	These results empirically confirm that choosing the initial SA state as a boundary point—achieved through quantum annealing—improves solution quality. This finding aligns with the earlier distributional analysis in \cref{sec:distr}, where we observed that most coordinates of optimal and near-optimal solutions lie on the boundary.\\
	
	Collectively, these analyses demonstrate that the QESA framework performs robustly across a wide range of QP scenarios. It achieves high-quality solutions for high-dimensional, ill-conditioned problems, outperforming classical solvers in several cases. Furthermore, key algorithmic choices—such as boundary-based initialization and direction optimization via QA—were shown to play a critical role in enabling this performance, despite using a relatively modest quantum iteration budget.

	\section{Conclusions}\label{sec:conclusion}
	
	In this work, we introduced \textit{Quantum Enhanced Simulated Annealing (QESA)}, a hybrid optimization framework that integrates quantum annealing (QA) into the classical simulated annealing (SA) algorithm. QESA applies quantum annealing to continuous quadratic optimization problems by using QA to guide search directions within a continuous domain. We demonstrated the utility of QESA on the important class of quadratic programming (QP) problems with box constraints, a setting that is both practically relevant and challenging due to its potential non-convexity and high dimensionality.
	
	Extensive experiments across a variety of problem sizes and matrix structures show that QESA consistently outperforms classical baselines, including standard simulated annealing (\SA) and the \texttt{SLSQP} solver from \texttt{scipy}. Compared to standalone QA applied to a discretized problem (\QA), QESA achieves significantly better accuracy and robustness while maintaining modest computational cost. In several cases, QESA even surpasses the solution quality of Gurobi when the latter is limited to a fixed runtime. These results demonstrate that quantum-guided exploration—particularly when used to determine search directions—can offer a significant advantage in hybrid optimization strategies.
	
	The scalability of QESA is limited by the capabilities of existing quantum annealing hardware. Constraints such as the number of available qubits, sparse connectivity, and embedding overheads impose limitations on the size of subproblems that can be efficiently encoded and solved on a quantum annealer. Looking ahead, future generations of quantum annealing hardware—with more qubits, better connectivity, and reduced noise—may enable QESA to tackle even more complex and high-dimensional continuous optimization problems.
	
	Several promising directions exist for extending QESA. These include exploring alternative quantum backends such as gate-based quantum processors or simulators, applying the framework to broader classes of continuous or mixed-integer problems, and improving the efficiency of classical–quantum interfacing to reduce communication overhead. As quantum annealing technology continues to evolve and mature, QESA holds strong potential as a versatile foundation for addressing increasingly complex continuous optimization tasks across diverse domains.

	\bibliographystyle{plain}
	\bibliography{quadratic_programming,quantum_annealing}

\begin{thebibliography}{10}

\bibitem{RevModPhys.90.015002}
Tameem Albash and Daniel~A. Lidar.
\newblock Adiabatic quantum computation.
\newblock {\em Rev. Mod. Phys.}, 90:015002, 2018.

\bibitem{arai2023effectiveness}
Shunta Arai, Hiroki Oshiyama, and Hidetoshi Nishimori.
\newblock Effectiveness of quantum annealing for continuous-variable
  optimization.
\newblock {\em Physical Review A}, 108(4):042403, 2023.

\bibitem{Barahona1982}
F.~Barahona.
\newblock On the computational complexity of {Ising} spin glass models.
\newblock {\em Journal of Physics A: Mathematical and General}, 15(10):3241,
  1982.

\bibitem{burer2009nonconvex}
Samuel Burer and Adam~N Letchford.
\newblock On nonconvex quadratic programming with box constraints.
\newblock {\em SIAM Journal on Optimization}, 20(2):1073--1089, 2009.

\bibitem{chang2020integer}
Chia~Cheng Chang, Chih-Chieh Chen, Christopher Koerber, Travis~S Humble, and
  Jim Ostrowski.
\newblock Integer programming from quantum annealing and open quantum systems.
\newblock {\em arXiv preprint arXiv:2009.11970}, 2020.

\bibitem{chapuis2017finding}
Guillaume Chapuis, Hristo Djidjev, Georg Hahn, and Guillaume Rizk.
\newblock Finding maximum cliques on a quantum annealer.
\newblock In {\em Proceedings of the Computing Frontiers Conference}, pages
  63--70, 2017.

\bibitem{choi2011minor}
Vicky Choi.
\newblock {Minor-embedding in adiabatic quantum computation: II.
  Minor-universal graph design}.
\newblock {\em Quantum Information Processing}, 10(3):343--353, 2011.

\bibitem{das2008colloquium}
Arnab Das and Bikas~K Chakrabarti.
\newblock Colloquium: Quantum annealing and analog quantum computation.
\newblock {\em Reviews of Modern Physics}, 80(3):1061, 2008.

\bibitem{DeAngelis1997}
Pasquale~L. De~Angelis, Panos~M. Pardalos, and Gerardo Toraldo.
\newblock {\em Quadratic Programming with Box Constraints}, pages 73--93.
\newblock Springer US, Boston, MA, 1997.

\bibitem{djidjev2024enhancing}
Hristo~N Djidjev.
\newblock Enhancing quantum annealing accuracy through replication-based error
  mitigation.
\newblock {\em Quantum Science and Technology}, 9(4):045034, 2024.

\bibitem{farhi2000quantum}
Edward Farhi, Jeffrey Goldstone, Sam Gutmann, and Michael Sipser.
\newblock Quantum computation by adiabatic evolution.
\newblock {\em arXiv preprint quant-ph/0001106}, 2000.

\bibitem{gurobi}
{Gurobi Optimization, LLC}.
\newblock {\em Gurobi Optimizer Reference Manual}, 2025.
\newblock \url{https://www.gurobi.com}, retrieved March 2025.

\bibitem{iftakher2023mixed}
Ashfaq Iftakher, Monzure-Khoda Kazi, and MM~Faruque Hasana.
\newblock Mixed-integer quadratic optimization using quantum computing for
  process applications.
\newblock {\em Proceeding of the Foundations of Computer Aided Process
  Operations/Chemical Process Control}, pages 1--6, 2023.

\bibitem{kadowaki1998quantum}
Tadashi Kadowaki and Hidetoshi Nishimori.
\newblock Quantum annealing in the transverse ising model.
\newblock {\em Phys. Rev. E}, 58:5355--5363, 1998.

\bibitem{king2022coherent}
Andrew~D. King, Sei Suzuki, Jack Raymond, et~al.
\newblock Coherent quantum annealing in a programmable 2,000-qubit {Ising}
  chain.
\newblock {\em Nature Physics}, 18(11):1324--1328, 2022.

\bibitem{kirkpatrick1983optimization}
Scott Kirkpatrick, C~Daniel Gelatt~Jr, and Mario~P Vecchi.
\newblock Optimization by simulated annealing.
\newblock {\em science}, 220(4598):671--680, 1983.

\bibitem{kwok2020graph}
Julia Kwok and Kristen Pudenz.
\newblock Graph coloring with quantum annealing.
\newblock {\em arXiv preprint arXiv:2012.04470}, 2020.

\bibitem{Lucas2014}
A.~Lucas.
\newblock {Ising formulations of many NP problems}.
\newblock {\em Front Physics}, 2(5), 2014.

\bibitem{lucas2014ising}
Andrew Lucas.
\newblock Ising formulations of many np problems.
\newblock {\em Frontiers in Physics}, 2:5, 2014.

\bibitem{Morita_2008}
Satoshi Morita and Hidetoshi Nishimori.
\newblock Mathematical foundation of quantum annealing.
\newblock {\em Journal of Mathematical Physics}, 49(12), 2008.

\bibitem{nocedal2006quadratic}
Jorge Nocedal and Stephen~J Wright.
\newblock Quadratic programming.
\newblock {\em Numerical optimization}, pages 448--492, 2006.

\bibitem{ottaviani2018lowranknonnegativematrix}
Daniele Ottaviani and Alfonso Amendola.
\newblock Low rank non-negative matrix factorization with d-wave 2000q, 2018.

\bibitem{pardalos1991quadratic}
Panos~M Pardalos and Stephen~A Vavasis.
\newblock Quadratic programming with one negative eigenvalue is np-hard.
\newblock {\em Journal of Global optimization}, 1(1):15--22, 1991.

\bibitem{pearson2019analog}
Adam Pearson, Anurag Mishra, Itay Hen, and Daniel~A Lidar.
\newblock Analog errors in quantum annealing: doom and hope.
\newblock {\em npj Quantum Information}, 5(1):107, 2019.

\bibitem{pelofske2019solving}
Elijah Pelofske, Georg Hahn, and Hristo Djidjev.
\newblock Solving large minimum vertex cover problems on a quantum annealer.
\newblock In {\em Proceedings of the 16th ACM International Conference on
  Computing Frontiers}, pages 76--84, 2019.

\bibitem{simanneal}
Matthew Perry.
\newblock Python module for simulated annealing.
\newblock \url{https://github.com/perrygeo/simanneal}.
\newblock Accessed: 2025-03-28.

\bibitem{pudenz2014error}
Kristen~L Pudenz, Tameem Albash, and Daniel~A Lidar.
\newblock Error-corrected quantum annealing with hundreds of qubits.
\newblock {\em Nature communications}, 5(1):1--10, 2014.

\bibitem{sahni1974computationally}
Sartaj Sahni.
\newblock Computationally related problems.
\newblock {\em SIAM Journal on computing}, 3(4):262--279, 1974.

\bibitem{D-Wave-ICE}
D-Wave Systems.
\newblock Error sources for problem representation.
\newblock \url{https://docs.dwavesys.com/docs/latest/c_qpu_ice.html}, 2025.
\newblock Accessed February 2025.

\bibitem{van1987simulated}
Peter J.~M. van Laarhoven and Emile H.~L. Aarts.
\newblock Simulated annealing.
\newblock In {\em Simulated Annealing: Theory and Applications}, pages 7--15.
  Springer Netherlands, Dordrecht, 1987.

\bibitem{2020SciPy-NMeth}
Pauli Virtanen, Ralf Gommers, et~al.
\newblock {SciPy} 1.0: Fundamental algorithms for scientific computing in
  python.
\newblock {\em Nature Methods}, 17:261--272, 2020.

\end{thebibliography}
	
\end{document}